\documentclass[aps,pra,superscriptaddress,showpacs,twocolumn]{revtex4-2}
\usepackage{bm}
\usepackage{amsmath}
\usepackage{amssymb}
\usepackage{slashed}
\usepackage{float}
\allowdisplaybreaks
\usepackage{subcaption}
\usepackage{dcolumn}
\usepackage{graphicx}
\newcolumntype{.}{D{x}{}{-1}}

\newcolumntype{w}[1]{D{.}{.}{#1}}
\newcolumntype{L}{>{$}l<{$}}

\newcommand{\Za}{{Z\alpha}}

\usepackage{graphicx}
\usepackage{xcolor}
\usepackage{nicefrac}

\newcommand{\lbr}{\langle}
\newcommand{\rbr}{\rangle}

\begin{document}

\title{Higher-order QED corrections to the hyperfine splitting in $\bm{^3}$He}

\author{Vojt\v{e}ch Patk\'o\v{s}}
\affiliation{
Charles University,  Ke Karlovu 3, 121 16 Prague
2, Czech Republic}

\author{Vladimir A. Yerokhin}
\affiliation{Max–Planck–Institut f\"ur Kernphysik, Saupfercheckweg 1, 69117 Heidelberg, Germany}
\affiliation{
University of Warsaw, Pasteura 5, 02-093 Warsaw, Poland}

\author{Krzysztof Pachucki}
\affiliation{
University of Warsaw, Pasteura 5, 02-093 Warsaw, Poland}

\begin{abstract}
We present a calculation of the hyperfine splitting of the $2^3S$ state in the $^3$He atom
with inclusion of all QED effects up to $\alpha^3E_F$, where $E_F$ is the Fermi splitting.
Using the experimental value of the $1S$ hyperfine splitting in $^3$He$^+$, we eliminate uncertainties from the nuclear structure
and obtain the theoretical prediction for $^3$He of $\nu_\mathrm{hfs}=
-6\,739\,701\,181(41)$~Hz, which is in perfect agreement
with the experimental value $-6\,739\,701\,177(16)$~Hz [S.~D.~Rosner and F.~M.~Pipkin, Phys.~Rev.~A
{\bf 1}, 571 (1970)].
This result constitutes a 40-fold improvement in precision as compared to the previous value and
is the most accurate theoretical prediction ever obtained for a non-hydrogenic system.

\end{abstract}
\maketitle

{\sl Introduction.---}
Interaction of the magnetic moment of the nucleus with that
of the electron leads to the splitting of atomic
energy levels known as the hyperfine splitting (hfs).
The hfs of atoms can be measured with outstanding accuracy,
e.g., the ground-state hfs of hydrogen is experimentally known up to
12 digits \cite{essen:71,essen:73,karshenboim:05:rep}. This makes hfs an excellent candidate
for high-precision tests of the quantum electrodynamics (QED)
of bound states \cite{eides:01} and for searches of physics beyond the standard model \cite{karshenboim:11:hfs}.

An impediment to performing such tests
is that theoretical hfs predictions are severely limited
by nuclear effects, which are manifested already at the $10^{-4}$ level and cannot be
accurately calculated at present. This impediment can be circumvented
\cite{sternheim:63,karshenboim:02:epjd,karshenboim:05:rep} by making use of the
fact that the hfs of different atomic states is strongly correlated, being
largely proportional to the electron charge density at the nucleus.
Therefore, one can employ an experimental hfs value measured for one state in order
to obtain an improved theoretical prediction for another state.
This idea has been realized for hydrogen
\cite{karshenboim:02:epjd,jentschura:06:hfs}, where theory
was able to predict the hfs of excited $nS$ states with a sub-Hertz accuracy
with help of the experimental $1S$ hfs value, in agreement
with the recent  measurement of $2S$ hfs \cite{bullis:23} which also achieved sub-Hertz accuracy.

The same idea has been recently applied to the HD$^+$ molecule. Specifically,
the nuclear-structure effects have been eliminated by using experimental hfs values
of H and D atoms.
The spin-averaged transitions measured by several groups
\cite{alighanbari_20, patra_20, kortunov_21}
agreed very well with theoretical predictions \cite{korobov:21:rov} and
provided the most accurate determination of the electron mass.
However, the hfs from one of these measurements \cite{patra_20} deviated by 9$\,\sigma$
from the theoretical predictions \cite{korobov_22}.
This disagreement is very intriguing, because HD$^+$ is a molecule with
only one electron and can be calculated almost as precisely as hydrogen atom.

Another system whose hfs can be accurately measured and predicted theoretically
is the helium atom. Up to now, its theoretical calculations were hampered by
severe difficulties in QED
treatment of the electron-electron correlations, which limited
the theoretical accuracy on the level
of about 1~kHz \cite{pachucki:01:jpb,pachucki:12:hehfs}.
In this Letter, we demonstrate that the rigorous QED treatment of hfs
of few-electron atoms
is possible up to the order of $\alpha^3E_F$, where $\alpha$ is the fine-structure
constant and $E_F$ is the Fermi splitting. We perform numerical calculations
for the $2^3S$ state of $^3$He and use the experimental He$^+$ hfs value
to eliminate nuclear uncertainties.
Our calculation increases the theoretical accuracy by more than an order of magnitude.
The updated theoretical result has an accuracy of 41~Hz and is in excellent
agreement with the experimental value \cite{rosner:70}.
This constitutes the strongest test of QED hfs theory in few-body
systems, which is of particular importance now in view
of the discrepancy observed in HD$^+$ \cite{patra_20}.

{\sl Hyperfine splitting.---}
The QED theory of hfs in the $S$ state starts with
the leading contribution given by the so-called Fermi splitting $E_F$,
\begin{align}\label{eq:0}
E_F \equiv \langle V_F \rangle = \frac{4\,\pi\,Z\alpha}{3\,m\,M}\,g\,
\langle \vec I\cdot [\vec s_1\,\delta^3(r_1) + \vec s_2\,\delta^3(r_2)]\rangle\,,
\end{align}
where $\vec{I}$ and $M$ are the nuclear spin and mass, respectively, $\vec{s}_i$ and
$m$ are the spin and the mass of the electrons, respectively,
$Z$ is the nuclear charge number,
$\alpha$ is the fine-structure constant, and
the natural nuclear $g$ factor is defined from the nuclear magnetic moment $\vec\mu$ by
$\vec\mu = Ze/(2M)\,g\,\vec I$.
The leading QED correction to the Fermi splitting is obtained by multiplying $E_F$ by the magnetic moment anomaly
of the free electron $\kappa = (g_e-2)/2$. Rigorous theory of the hfs of light atomic systems is constructed
within the nonrelativistic quantum electrodynamics (NRQED) in the form of an expansion
in the fine-structure constant.
We represent it as follows
\begin{equation}\label{eq:1}
E_\mathrm{hfs} = E_F\,(1+\kappa) + E^{(6)} + E^{(7)} + E^{(8)} + E_\mathrm{nuc} + E_\mathrm{rec}\,.
\end{equation}
Here,  $E^{(n)}$ are the QED effects of order $m\alpha^n$ for the point-like and infinitely-heavy nucleus,
$E_\mathrm{nuc}$ represents the nuclear structure effects,
and $E_\mathrm{rec}$ is nuclear recoil correction.
The nuclear effects $E_\mathrm{nuc}$
cannot be calculated accurately at present, so we extract them from the experimental hfs value in He$^+$.
Calculations of the leading hfs term (i.e., $E_F$) are well established at present \cite{morton:06}.
QED effects of order $\alpha^2E_F$ (i.e., $E^{(6)}$) were calculated
in Refs.~\cite{pachucki:01:jpb,pachucki:12:hehfs}. Here we calculate the QED effects of order
$\alpha^3E_F$ (i.e., $E^{(7)}$) and
the dominant part of the recoil correction of order $\alpha^2\,(m/M)\,E_F$ (i.e., $E_\mathrm{rec}$),
which leads to a drastic improvement of theoretical accuracy.

{\sl QED effects of order $\alpha^3\,E_F$.---}
The derivation described in Appendix provides
the complete expression for the $m\alpha^7$ QED correction
which does not contain any divergences and can be used for numerical evaluation.
The final result is separated into the low-energy $(E_L)$, the
first-order matrix-elements $(E_{\rm fo})$ and the second-order matrix-elements
$(E_{\rm sec})$ parts. The first-order and second-order contributions are further split
into the self-energy (se) and vacuum-polarization (vp) parts. We thus write,
in atomic units and with the prefactor $m\alpha^7$ pulled out, $E^{(7)} \equiv m\alpha^7{\cal E}^{(7)}$,
\begin{align}
{\cal E}^{(7)}_{\rm hfs} = &\ {\cal E}_L + {\cal E}_{\rm fo}\mathrm{(se)} + {\cal E}_{\rm sec}\mathrm{(se)} + {\cal E}_{\rm fo}\mathrm{(vp)} + {\cal E}_{\rm sec}\mathrm{(vp)}\,.
\end{align}
The low-energy Bethe-logarithm-type correction ${\cal E}_{L}$ is defined by Eq.~(\ref{38}).
The first-order contributions
${\cal E}_{\rm fo}$ can be conveniently expressed in terms of expectation values of
$Q_i$ operators, which were encountered in our previous investigation
of the Lamb shift \cite{patkos:21:helamb} and are defined in Table \ref{oprsQ1}.
The result for the self-energy is
\begin{widetext}
\begin{align}\label{fofinQ}
{\cal E}_{\rm fo}\mathrm{(se)}
  = &\  \frac{g\,m}{2\,\pi\,M}\,\langle\vec I\cdot\vec S\rangle\,\bigg\{
  \frac{1}{9}\, \bigg(\frac{71}{3}+32\ln\frac{\alpha^{-2}}{2}\bigg)\, Z^2\, Q_1\,Q_{53}
 +\bigg(\frac{143}{108} + \frac89\ln\frac{\alpha^{-2}}{2}\bigg)\,Z^2\, Q_{57}
 \nonumber\\
&\ -\frac{1}{3}\, \bigg(\frac{85}{6}+16\ln\frac{\alpha^{-2}}{2} \bigg) \,\frac{Z^2}{2}\, Q_3
 - \frac{56}{9}\,Z\,Q_9\,Q_{53}
 +\frac{56}{9}\,Z\, Q_{59}
-\frac{13}{12}\,Z\,Q_{18}
 + \frac{4Z}{3}\, E^{(4)}\,Q_{53}
\nonumber\\&\
 +\frac{2Z}{3} \Big(
-2 E_0 Q_{13} + Q_{17} + E_0^2 Q_{53} + 2 Z E_0 Q_{11} + 2 Z E_0 Q_{12} - 2 Z Q_{14} - 2 Z Q_{16} + 3 Z^2 Q_{15} +  Z^2 Q_{56}
 \Big)
\nonumber\\&\
 -\frac{Z}{3}\,Q_{28}
 + \frac{2Z}{3}\,Q_{24}
+\frac{Z}{36}\bigg(\frac{77}{6} + 16 \ln\frac{\alpha^{-2}}{2} \bigg)\,Q_{51}
-\frac{Z}{36}\bigg(\frac{95}{3} + 32 \ln\frac{\alpha^{-2}}{2} \bigg)
\Big( E_0\,Q_1 - Q_3 -\frac12 \,Q_4 \Big)
\nonumber\\
&\ + \bigg[-\frac76 - \frac{44\pi^2}{27} - \frac{10}{3}\zeta(3) + \frac{896}{27}\ln 2+ \frac{16}{9}\ln^2 2
- \frac{938}{27}\ln\alpha - \frac{64}{9}\ln^2\alpha + \frac{256}{9}\ln 2\ln\alpha \bigg]
\,\frac{Z^3}{4}\,Q_1
\bigg\} \,.
\end{align}
\end{widetext}
The second-order self-energy contribution is given by
\begin{align}\label{sofin2}
{\cal E}_{\rm sec}\mathrm{(se)} = &\ \frac{g\,m}{2\,\pi\,M}\,\langle\vec I\cdot\vec S\rangle\,\bigg\{
\frac{2}{9}\,
\bigg[\bigg(\frac{5}{6}+\ln\frac{\alpha^{-2}}{2}\bigg)\,S_1 - 7\,S_2
\nonumber \\ &\
+ \frac32\, S_3\bigg] + \frac{Z}{3}\,
\bigg(\frac{Z}{2}\,S_4 - S_5\bigg)
 -\frac{Z}{8}\,S_6
\bigg\}\,,
\end{align}
where the second-order matrix elements $S_i$ are defined in Table \ref{table:so}.
For the vacuum-polarization we obtain the following results
\begin{align}\label{finvp2}
{\cal E}_{\rm fo}\mathrm{(vp)} = &\
-\frac{g\,m}{45\,\pi\,M}\,\langle\vec I\cdot\vec S\rangle\,
\bigg[
16 Z^2 \, Q_1\,Q_{53}
+ 2 Z\, Q_{51}
 \nonumber\\ &
+4Z(1-3Z)\, Q_3
- 4 Z\, E_0\,Q_1 +2Z\, Q_4
\nonumber
\\ &\
+4Z^2 Q_{57}
+ Z^3\,\Big(\frac{236}{15} + 8\ln\alpha\Big)\, Q_1
\bigg]\,.
\end{align}
and
\begin{align}\label{finvpsec}
{\cal E}_{\rm sec}\mathrm{(vp)} = &\
-\frac{g\,m}{45\,\pi\,M}\,\langle\vec I\cdot\vec S\rangle\,S_1\,.
\end{align}

The numerical calculations of the $m\alpha^7$ corrections are carried out
with the basis set of exponential functions
$e^{-\alpha_i\,r_1-\beta_i\,r_2-\gamma_i\,r}$ introduced by Korobov \cite{korobov:00}, where $r = |\vec r_1-\vec r_2|$.
The method of calculations follows the one developed in our previous investigations
and described in Ref.~\cite{yerokhin:21:hereview}. The calculation of the low-energy
Bethe-logarithm-type
contribution follows our previous work~\cite{yerokhin:18:betherel}.
Numerical results for the individual $m\alpha^7$ corrections to the hfs of the $2^3S$ state
in $^3$He are presented in Table~\ref{table:E7}.

\begin{table}
\caption{First-order matrix elements for the $2^3S$ state, numerical results are from Ref. \cite{patkos:21:helamb} }
\label{oprsQ1}
\begin{ruledtabular}
\begin{tabular}{ld}
Operator  &  \langle Q_i\rangle\\ \hline
$Q_1 =4 \pi \delta^3 (r_1)$   				     &  16.592\,071         \\
$Q_3 =4 \pi \delta^3(r_1)/r_2$                  	     &   4.648\,724     \\	
$Q_4 =4 \pi \delta^3(r_1)\, p_2^2$ 	                     &   2.095\,714     \\
$Q_9 =1/r^3$                    	                     &   0.038\,861   \\
$Q_{11}=1/r_1^2$                	                     &   4.170\,446     \\
$Q_{12}=1/(r_1 r_2)$            	                     &   0.560\,730      \\
$Q_{13}=1/(r_1 r)$              	                     &   0.322\,696     \\
$Q_{14}=1/(r_1 r_2 r)$          	                     &   0.186\,586    \\
$Q_{15}=1/(r_1^2 r_2)$					     &   1.242\,704     \\
$Q_{16}=1/(r_1^2 r)$					     &   1.164\,599     \\
$Q_{17}=1/(r_1 r^2)$   					     &   0.112\,360     \\
$Q_{18}=(\vec{r}_1\cdot\vec r)/(r_1^3 r^3)$                  &   0.011\,331     \\
$Q_{24}=p_1^i\,(r^i r^j+\delta^{ij} r^2)/(r_1 r^3)\, p_2^j$  &   0.002\,750     \\
$Q_{28}=p_1^2\,/r_1\, p_2^2$				     &   1.597\,727     \\
$Q_{51} =4 \pi\,\vec p_1\,\delta^3(r_1)\,\vec p_1 $	         &  0.009\,993     \\
$Q_{53} =1/r_1$						                         &  1.154\,664  \\
$Q_{56}=1/r_1^3$   					                         &-23.022\,535  \\
$Q_{57}=1/r_1^4$   					                         & 25.511\,837  \\
$Q_{59}=1/(r_1 r^3)$				                         &  0.051\,914   \\
\end{tabular}
\end{ruledtabular}
\end{table}

\begin{table}
\caption{Second-order corrections for the $2^3S$ state.}
\label{table:so}
\begin{ruledtabular}
\begin{tabular}{lw{5.6}}
\multicolumn{1}{l}{Term} &
        \multicolumn{1}{c}{Value}
\\
\hline\\[-5pt]
$S_1 = \Big<V_R\,\frac{1}{(E_0-H_0)'}\,V_R \Big>$            & -2634.595\,12 \\
$S_2 = \Big<V_R\,\frac{1}{(E_0-H_0)'}\,\frac{1}{r^3}\Big>$   &     0.371\,13 \\
$S_3 = \Big<V_R\,\frac{1}{(E_0-H_0)'}\,H_R \Big>$            &   202.676\,07 \\
$S_4 = \bigg\langle\Big(\frac{\vec r_1}{r_1^3}\times\vec p_1+\frac{\vec r_2}{r_2^3}\times\vec p_2\Big)\,$   \\
\hspace{3ex}$\times\frac{1}{(E_0-H_0)'}\,
\Big(\frac{\vec{ r}_1}{r_1^3}\times\vec{ p}_1+\frac{\vec{ r}_2}{r_2^3}\times\vec{ p}_2\Big)\Big> $      &    -0.004\,69   \\
$ S_5 = \bigg\langle\Big(\frac{\vec r_1}{r_1^3}\times\vec p_1+\frac{\vec r_2}{r_2^3}\times\vec p_2\Big)$    \\
\hspace{3ex}$\times\frac{1}{(E_0-H_0)'}\,
\frac{\vec{ r}}{r^3}\times(\vec{ p}_1 - \vec{ p}_2)\Big> $        &    -0.007\,07 \\
$S_6 = \Big< \Big(\frac{\delta^{ij}}{r_1^3} - \frac{3 r_1^i r_1^j}{r_1^5}
+\frac{\delta^{ij}}{r_2^3} - \frac{3 r_2^i r_2^j}{r_2^5}\Big) $    \\
\hspace{3ex}$\times\frac{1}{(E_0-H_0)'}\,\left(
\frac{\delta^{ij}}{r^3}-3\,\frac{r^i r^j}{r^5}\right)\Big>$  &    -0.01128 \\
\end{tabular}
\end{ruledtabular}
\end{table}

\begin{table}
\caption{$m\alpha^7$ corrections to the hfs of the $2^3S$ state.
$\cal E$ are in units of $\alpha^3\,E_F$ and
$\delta^{(3)} = {\cal E}^{(7)} \alpha^3$.
}
\label{table:E7}
\begin{ruledtabular}
\begin{tabular}{lw{-5.3}}
\multicolumn{1}{l}{Term} &
        \multicolumn{1}{c}{Value}
\\
\hline\\[-5pt]
 ${\cal E}_L$  &     22.05873(88) \\
 ${\cal E}_\mathrm{fo}$(se)   &       8.31316 \\
 ${\cal E}_\mathrm{sec}$(se)  &    -83.11218 \\
 ${\cal E}_\mathrm{fo}$(vp)   &      0.88943 \\
 ${\cal E}_\mathrm{sec}$(vp)  &      1.68478 \\[1ex]
 ${\cal E}^{(7)}(\mathrm{He})$&    -50.16609(88) \\
 ${\cal E}^{(7)}(\mathrm{He}^+)$&  -50.64036 \\[1ex]
 ${\cal E}^{(7)}(\mbox{\rm He-He}^+)$   &      0.47428(88) \\
 $\delta^{(3)}(\mbox{\rm He-He}^+)$ &  0.1843(3)\times 10^{-6}
\end{tabular}
\end{ruledtabular}
\end{table}

{\sl Hyperfine mixing correction.---}
For the $2^3S_1$ state the nuclear recoil effects are dominated
by the second-order hyperfine correction
induced by the Fermi contact interaction $V_F$,
specifically, by the $2^3S_1$-$2^1S_0$ mixing contribution.
The Fermi interaction mixes states with different
values of the total momentum $J$ and the
$2^3S_1$-$2^1S_0$ mixing is strongly enhanced
because of the small energy difference of these states \cite{pachucki:01:jpb}.
The leading mixing contribution is of order
$\alpha^2 \,(m/M)\,E_F$ and given by
\begin{align}
E_{\rm mix}^{(6)} =&\ \frac{\bigl\langle 2^3S| V_F | 2^1S\bigr\rangle^2}{E_0(2^3S)-E_0(2^1S)} \,,
\label{49}
\end{align}
which leads to a surprisingly large result,
$E_{\rm mix}^{(6)} = - 8.992\,1\times 10^{-6}\,E_F\,.$
The numerical value of $E_{\rm mix}^{(6)}$ is so large that we have to consider
higher-order corrections to it, which are small but not negligible at our level
of interest. First, we consider the recoil correction to Eq.~(\ref{49}). Using
the matrix element with full mass dependence
\begin{align}
4\,\pi\,\langle 2^3S_1|\big[\delta^3(r_1) - \delta^3(r_2)\big]|2^1S_0\rangle_{M}
  =& \Bigl(\frac{\mu}{m}\Bigr)^3\,29.135\,080\,,
\end{align}
(with $\mu = mM/(m+M)$) and including the recoil correction in the energy denominator, we obtain
$\delta E_\mathrm{mix,rec} = 0.003\,2\times 10^{-6}\,E_F$ for the nuclear mass correction beyond that in $E_F$\,.
Second, we take into account the corrections due to the anomalous magnetic moment and the nuclear
effects to the operator and the relativistic correction to the energies,
\begin{align}
 \delta E_\mathrm{mix,rad} =&\  E_{\rm mix}^{(6)}\,
    \biggl[ \Big(1+\kappa+\frac{E_\mathrm{nuc}}{E_F}\Big)^2 -1
 -\frac{\delta E_\mathrm{rel}}{\delta E} \biggr]\,,
 \end{align}
where $\delta E_\mathrm{rel}/\delta E$ is the relative contribution of the relativistic
correction to the $2^3S$-$2^1S$ energy difference. This
yields $\delta E_\mathrm{mix,rad} = -0.015\,2\times 10^{-6}\,E_F$\,.
Finally, we consider the correction due to the mixing with higher excited states.
The summation over the complete spectrum in the second-order contribution
will lead to the infinite result, which indicates that
it is not a complete recoil correction. Following
Ref.~\cite{pachucki:01:jpb}, we here consider the normalized difference of this correction
between helium atom and helium ion,
\begin{align}
\delta E_{\rm mix,exc} &\ = \langle V_F\,\frac{1}{(E-H)'}\,V_F\rangle\bigg|_\mathrm{He}
\nonumber \\ &
 \!\!\!\!\!\!\!\!\!\!\!\!\!\!
-\frac{3}{4}\,\frac{\langle\pi\,(\delta^3(r_1)+\delta^3(r_2))\rangle}{8}
\langle V_F\,\frac{1}{(E-H)'}\,V_F\rangle\bigg|_\mathrm{He^+}\,,
\end{align}
which is finite and yields a numerical contribution of
$\delta E_{\rm mix,exc} = 0.010\,3\cdot 10^{-6}\,E_F$\,.
Finally, the total recoil correction is given by the sum
$\Delta E_{\rm rec} = E_{\rm mix}^{(6)} + \delta E_\mathrm{mix,rec} + \delta E_\mathrm{mix,rad} +
\delta E_\mathrm{mix,exc}$, with the numerical result presented in
Table~\ref{table:total}.

{\sl Results and discussion.---}
For the final analysis
it is convenient to represent all corrections to hfs as multiplicative factors to $E_F$,
\begin{align}\label{eq:22}
E_\mathrm{hfs} =&\ E_F\,(1 + \delta)\,,
\end{align}
where
\begin{align}
 \delta =&\kappa
+  \delta^{(2)}
+ \delta^{(3)}+ \delta^{(4)}
+ \delta_\mathrm{nuc} + \delta_\mathrm{rec} \,,
\end{align}
which is equivalent to Eq.~(\ref{eq:1}) with $\delta^{(k)} = E^{(k+4)}/E_F$.
The main advantage of this representation is that the
$\delta$ coefficients are strongly correlated with those in He$^+$.
In order to exploit this correlation, we split $\delta$ in Eq.~(\ref{eq:22})
into two parts,
\begin{align}
\delta(\mathrm{He}) =&\
\delta(\mathrm{He}^+) + \delta(\mbox{\rm He-He}^+)\,,
\end{align}
where $\delta(\mathrm{He}^+)$ will be extracted from the experiment on He$^+$ and
$\delta(\mbox{\rm He-He}^+)$ is calculated theoretically.

The individual theoretical contributions to $ \delta(\mbox{\rm He-He}^+)$ are presented
in Table~\ref{table:total}.
The leading term, $\delta^{(2)}$, is of order $\alpha^2E_F$.
It was calculated first by one of the authors in Ref.~\cite{pachucki:01:jpb} and later
improved in Ref.~\cite{pachucki:12:hehfs}. The next-order QED correction of order $\alpha^3E_F$,
$\delta^{(3)}$, and the recoil contribution,
$\delta_\mathrm{rec}$, are calculated as described above.

In order to estimate the higher-order QED contribution $\delta^{(4)}(\mbox{\rm He-He}^+)$,
for which no direct calculations exist so far, we use results obtained in
Ref.~\cite{karshenboim:02:epjd} for the normalized difference of the hfs intervals in He$^+$,
$D_{21}= 8\,E_\mathrm{hfs}(2S)- E_\mathrm{hfs}(1S)$.
Specifically, we assume the ratio $\delta^{(4)}/\delta^{(3)}$ for the He-He$^+$ difference
to be the same as the corresponding ratio for $D_{21}$, with a 100\% uncertainty.
Similarly, we obtain the uncertainty of $\delta_\mathrm{rec}$ by
examining the ratio of $\delta_\mathrm{rec}^{(2+)}/\delta^{(2)}$ for $D_{21}$ and assuming the same
ratio holds for He-He$^+$ difference, thus obtaining the estimate of the omitted non-mixing hfs recoil contributions.

Adding the contribution $\delta(\mathrm{He}^+)$ inferred from the experimental result of
the $1S$ hfs in $^3$He$^+$ from Ref.~\cite{schneider:22},
we obtain the theoretical prediction for the He$(2^3S_1)$ hfs
with an accuracy of 41~Hz, see Table \ref{table:total},
in perfect agreement with the experimental result of Ref.~\cite{rosner:70}.

\begin{table}
\caption{Contributions to the $2^3S_1$ hfs of $^3$He.}
\label{table:total}
\begin{ruledtabular}
\begin{tabular}{lw{3.10}w{10.5}w{5.2}}
\multicolumn{1}{l}{Term} &
        \multicolumn{1}{c}{$\times 10^{-6}$} &
            \multicolumn{1}{c}{ [Hz]}&
            	  \multicolumn{1}{c}{$D_{21}$\,[kHz]}
\\
\hline\\[-5pt]
  $\delta^{(2)}(\mbox{\rm He-He}^+)$                 & 3.012\,0        &  -20\,279.     & -1\,152.44 \\
  $\delta^{(2+)}_\mathrm{rec}(\mbox{\rm He-He}^+)$   & -8.993\,7\,(21) &   60\,552.(14)  &  -0.80   \\
  $\delta^{(3)}(\mbox{\rm He-He}^+)$                 &0.184\,3(3)         &   -1\,241.(2)   & -36.03  \\
  $\delta^{(4)}(\mbox{\rm He-He}^+)$                 &0.005\,8\,(58)   &       -39.(39)  & -1.14    \\[1ex]
  $\delta(\mbox{\rm He-He}^+)$                       & -5.791\,6\,(62) &   38\,993.(41)    \\
  $1+ \delta(\mathrm{He}^+) $ \cite{schneider:22}    &                 &  -6\,739\,740\,174.  \\[1ex]
 $\nu_\mathrm{hfs,theo}(\mathrm{He})$                &                 &  -6\,739\,701\,181.(41)  \\
  $\nu_\mathrm{hfs,exp}(\mathrm{He})$ \cite{rosner:70} &               &  -6\,739\,701\,177.(16)  \\
\end{tabular}
\end{ruledtabular}
\end{table}

{\sl Conclusion.---}
In this Letter, we have demonstrated that advanced QED calculations are now capable of
predicting the hfs of helium with precision of several tens of Hz by using
the experimental hfs value for the corresponding hydrogen-like ion.
We derived formulas and performed numerical calculations for
the $2^3S$ state in $^3$He. This
improved the theoretical accuracy by a factor of 40 as compared to previous
calculations. The present theoretical precision of ${}^3\mathrm{He}(2^3S_1)$ hfs is 41~Hz, which makes it
the most accurate theoretical prediction ever achieved for non-hydrogenic systems.

The excellent agreement of theory and experiment for the helium hfs  contrasts sharply
with the $9\,\sigma$ discrepancy observed
for the HD$^+$ \cite{patra_20, korobov_22}.
The disagreement is very surprising, taking into account
the fact that the same theoretical approach is used in both systems.
If the discrepancy is confirmed in forthcoming studies, this would be a signal of some unknown physics.

Our calculations can also be extended to helium- and lithium-like ions,
in particular, to Li$^+$, for which accurate experimental results are available
\cite{clarke:03,sun:23}. The developed method can be used for extending the
advanced tests of QED to more complicated systems or,
alternatively, for determining the effective Zemach radii $\tilde r_Z$ of light nuclei.
The later direction is of particular interest in view of
the confirmed anomalies for the Zemach radii in $^6$Li and $^7$Li
\cite{puchalski:13, sun:23}
and a significant discrepancy for hfs in $\mu$D \cite{kalinowski:18}.

{\sl Acknowledgments.---}
K.P. and V.P. acknowledge support from the National Science Center (Poland) Grant No. 2017/27/B/ST2/02459.\\[1ex]

{\sl Appendix on derivation of the $\alpha^3 \,E_F$ effects.---}
The QED effects to hfs of the order $m\,\alpha^7$ ($= \alpha^3 \,E_F$) can be represented as
\begin{align}\label{eq:3}
E^{(7)} = &\  E_L + 2\,\langle H^{(4)}_{\rm hfs}\frac{1}{(E_0-H_0)'}\,H^{(5)}\rangle\nonumber\\
&\ +2\,\langle H^{(5)}_{\rm hfs}\frac{1}{(E_0-H_0)'}\,H^{(4)}\rangle + \langle H^{(7)}_{\rm hfs}\rangle\,.
\end{align}
Here, $E_L$ is the Bethe-logarithm-type low-energy contribution,
$H_0$ and $E_0$ denote the nonrelativistic Hamiltonian and its reference-state eigenvalue, respectively,
$H^{(4)}$ is the Breit-Pauli Hamiltonian of order
$m\alpha^4$, $H^{(5)}$ is the effective QED Hamiltonian of order $m\alpha^5$, and $H^{(4)}_{\rm hfs}$
and $H^{(5)}_{\rm hfs}$ are effective hfs Hamiltonians of order $m\alpha^4$ and $m\alpha^5$, respectively.
The Breit-Pauli Hamiltonian $H^{(4)}$ is well-known and given, e.g., by Eq.~(7) of Ref.~\cite{yerokhin:10:helike}.
The effective hfs Hamiltonian of order $\alpha^4$, $H^{(4)}_{\rm hfs}$, is responsible for the
leading-order hfs splitting. It can be obtained from Eqs.~(5)-(11) of Ref.~\cite{pachucki:12:hehfs}
by setting the electron magnetic anomaly to zero. The next-order effective hfs Hamiltonian
$H^{(5)}_{\rm hfs}$ is obtained from the same equations by picking up the linear part
in the electron magnetic anomaly.
The QED Hamiltonian $H^{(5)}$ is expressed as
\begin{align}\label{H5}
H^{(5)} = &\ \bigg(\frac{5}{6} -\frac{1}{5} + \ln\frac{\alpha^{-2}}{2\lambda}\bigg)\,\frac{4\,\alpha^2 Z}{3\,m^2}\,\big[\delta^3(r_1)+\delta^3(r_2)\big]
\nonumber \\ &\
-\frac{7\,\alpha^2}{3\,\pi\,m^2}\,\frac{1}{r^3} + H^{(5)}_{\rm fs}\,,
\end{align}
where $H^{(5)}_{\rm fs}$ is the spin-dependent part of $H^{(5)}$
and is given by Eq.~(14) of Ref.~\cite{yerokhin:10:helike}, and
$\lambda$ is the low-energy photon-momenta cutoff.
The dependence on the cutoff cancels out when all terms in Eq.~(\ref{eq:3}) are considered together,
which is explicitly demonstrated in the detailed derivation 
\cite{pachucki:23:arxiv}. Therefore, for simplicity, we will set $\lambda=1$ in the following formulas.

The low-energy Bethe-logarithm-type contribution $E_L$ comes from the virtual
photon momenta of the order $k\approx m\alpha^2$. It can be
represented (in atomic units, with the $m\alpha^7$ prefactor pulled out)
as
\begin{align} \label{38}
{\cal E}_{L} &\ = -\frac{2}{3\,\pi} \,
 \delta_{V_F}\, \Big< \vec{P}\,(H_0-E_0)\, \ln (H_0-E_0) \,
 \vec{P} \Big>
\,,
\end{align}
where $\delta_{V_F}\lbr S\rbr$ denotes the first-order perturbation of the matrix element $\lbr S\rbr$
by the Fermi contact interaction $V_F$ defined by Eq.~(\ref{eq:0}) and
$\vec{P} = \vec{p}_1+\vec{p}_2$ is the electron momentum operator.
The low-energy contribution $E_L$ is very similar to the Bethe-logarithm-type contribution $E_{L1}$
encountered in our previous study of the $m\alpha^7$ effects in the Lamb shift \cite{yerokhin:18:betherel}. In fact, all
necessary formulas for $E_L$ can be obtained by repeating the derivation of Ref.~\cite{yerokhin:18:betherel} for the
perturbation $V_F$ instead of the spin-independent Breit Hamiltonian. We thus refer the reader to our previous work
for detailed description of the evaluation of the low-energy contribution.

The second-order matrix elements in Eq.~(\ref{eq:3}) are problematic because of divergences originating from the
summation over the intermediate states. They arise when operators on the left
and on the right of the resolvent $1/(E_0-H_0)'$ are nearly singular so that their first-order
matrix elements are finite but the second-order matrix elements diverge. Specifically, there
are two such ``problematic'' operators in our case, the electron-nucleus Dirac $\delta$ function and
the spin-independent part of the Breit Hamiltonian $H^{(4)}_\mathrm{nfs}$ given by Eq.~(6) of
Ref.~\cite{pachucki:09:hefs}. In order to make
the divergences more tractable, we transfer them
to first-order matrix elements. This can be accomplished \cite{pachucki:06:hesinglet} by representing the problematic
operators as an anticommutator with the Schr\"odinger Hamiltonian $H_0$ plus some more regular operator.
Specifically,
\begin{align}
 4\pi Z\,&\,\big[\delta^3(r_1) + \delta^3(r_2)\big] = 2\,\Big\{ H_0-E_0,\frac{Z}{r_1}+\frac{Z}{r_2}\Big\} + V_R\,,\label{33}\\
&\, H^{(4)}_\mathrm{nfs} = -\frac{1}{4}\, \Big\{H_0-E_0,\frac{Z}{r_1}+\frac{Z}{r_2} \Big\} + H_R\,. \label{34}
\end{align}
The regularized operators $V_R$ and $H_R$  are acting on the eigenfunction of $H_0$ as
\begin{align}\label{VR}
V_R |\phi\rangle = -2\,Z\,\bigg(\frac{\vec r_1}{r_1^3}\cdot\vec\nabla_1+\frac{\vec r_2}{r_2^3}\cdot\vec\nabla_2\bigg)|\phi\rangle\,,
\end{align}
and
\begin{align}\label{HR}
&\ H_R|\phi\rangle = \bigg[\frac14 p_1^2 p_2^2 - \frac12 (E_0-V)^2 - \frac{1}{2}\,p_1^i\bigg(\frac{\delta^{ij}}{r}
+ \frac{ r^i r^j}{r^3}\bigg)\,p_2^j \nonumber\\
&\ - \frac{Z}{4}\frac{\vec r_1\cdot\vec\nabla_1}{r_1^3} - \frac{Z}{4}\frac{\vec r_2\cdot\vec\nabla_2}{r_2^3}
+\frac{1}{2}\frac{\vec r}{r^3}\cdot(\vec\nabla_1-\vec\nabla_2)\bigg]|\phi\rangle\,,
\end{align}
with $V = -Z/r_1-Z/r_2+ 1/r$.
The second-order contribution is thus transformed into
\begin{align}
2\,\langle H^{(4)}_{\rm hfs}&\, \frac{1}{(E_0-H_0)'}\,H^{(5)}\rangle
 +2\,\langle H^{(5)}_{\rm hfs}\frac{1}{(E_0-H_0)'}\,H^{(4)}\rangle
 \nonumber \\ &
= m\alpha^7\big[ {\cal E}_{\sec}(\mathrm{se})+ {\cal E}_{\sec}(\mathrm{vp}) + {\cal E}_{\mathrm{fo},A}\big]\,,
\end{align}
where ${\cal E}_{\sec}(\mathrm{se})$ and $ {\cal E}_{\sec}(\mathrm{vp})$
are regularized second-order corrections given by Eqs.~(\ref{sofin2}) and (\ref{finvpsec}), and
${\cal E}_{\mathrm{fo},A}$ is the first-order contribution given by
\begin{widetext}
\begin{align}\label{foA}
{\cal E}_{\mathrm{fo},A}
= &\ \langle\vec I\cdot\vec S\rangle\,\frac{g\,m}{3\,\pi\,M}\bigg\{
 \frac{1}{3}\,\bigg[
 \bigg(\frac{5}{6}-\frac15+\ln\frac{\alpha^{-2}}{2}\bigg)\bigg(\big\langle 16\pi \,Z[\delta^3(r_1)+\delta^3(r_2)]\big\rangle\bigg\langle\frac{Z}{r_1}+\frac{Z}{r_2}\bigg\rangle
-\bigg\langle 16\pi\,Z[\delta^3(r_1)+\delta^3(r_2)]\nonumber\\
&\ \times\bigg(\frac{Z}{r_1}+\frac{Z}{r_2}\bigg)\bigg\rangle +2\bigg\langle\frac{Z^2}{r_1^4}+\frac{Z^2}{r_2^4}\bigg\rangle\bigg)
 - 14\bigg\langle\frac{1}{r^3}\bigg\rangle\bigg\langle\frac{Z}{r_1}+\frac{Z}{r_2}\bigg\rangle
+14\bigg\langle\frac{1}{r^3}\bigg(\frac{Z}{r_1}+\frac{Z}{r_2}\bigg)\bigg\rangle\bigg]\nonumber\\
&\ + \frac{1}{2}\bigg[
\frac{1}{4}\bigg\langle\frac{Z^2}{r_1^4}+\frac{Z^2}{r_2^4}
-2\bigg(\frac{Z\vec r_1}{r^3_1}-\frac{Z\vec r_2}{r^3_2}\bigg)\cdot\frac{\vec r}{r^3}\bigg\rangle
+\bigg\langle\bigg(\frac{Z}{r_1}+\frac{Z}{r_2}\bigg) (E_0-V)^2\bigg\rangle
-\frac{1}{2}\bigg\langle p_1^2\bigg(\frac{Z}{r_1}+\frac{Z}{r_2}\bigg)p_2^2\bigg\rangle\nonumber\\
&\ + 2\, E^{(4)}\bigg\langle\frac{Z}{r_1}+\frac{Z}{r_2}\bigg\rangle
+ \bigg\langle p_1^i\bigg(\frac{Z}{r_1}+\frac{Z}{r_2}\bigg)\bigg(\frac{\delta^{ij}}{r}+\frac{r^i r^j}{r^3}\bigg)p_2^j\bigg\rangle
 -\big\langle \pi\,Z\,\big[\delta^3(r_1)+\delta^3(r_2)\big]\big\rangle\bigg\langle\frac{Z}{r_1}+\frac{Z}{r_2}\bigg\rangle\bigg]\bigg\}\,,
\end{align}
\end{widetext}
where $m\alpha^4 E^{(4)}$ is the relativistic correction to the energy centroid.
The singularities are now moved into the first-order terms in Eq.~(\ref{foA}).
Divergencies in singular operators $Z^2/r_a^4$ and $Z^3/r_a^3$ are handled according to Ref.~\cite{patkos:21:helamb}.

$H^{(7)}_\mathrm{hfs}$  is an effective Hamiltonian of order $m\, \alpha^7$.
It comes from the one-loop self-energy and the one-loop vacuum polarization only,
because no photon-exchange terms contribute at this order. It is represented as
\begin{align}\label{H7se}
H^{(7)}_{\rm hfs} = &\ H^{(7)}_{{\rm hfs},A} + H^{(7)}_{{\rm hfs},B}
  + \ldots\,,
\end{align}
where $\ldots$ denotes terms that are proportional to the electron-nucleus Dirac $\delta$ function, $\propto Z^3\,\delta^3(r_a)$.
At the current stage of the derivation we drop such terms; the corresponding
contribution will be restored later by matching the high-$Z$ limit of the
obtained formulas to the known hydrogenic result.
$H^{(7)}_{{\rm hfs},A}$ is induced by the spin-dependent terms in
the generalized Breit-Pauli Hamiltonian $H_\mathrm{BP}$ (see Eqs.~(15)-(17) of Ref.~\cite{pachucki:04:lwqed})
that are proportional to the magnetic moment anomaly,
\begin{align}\label{H7a}
H^{(7)}_{ {\rm hfs},A} =&\
\kappa
\sum_a \bigg[\frac{Z\alpha}{2\,m^2}\, \vec\sigma_a\cdot \frac{\vec r_a}{r_a^3}\times \big[-e\vec A_a\big]\\
&\ - \frac{e}{16\,m^3} \, \vec\sigma_a\cdot\Delta\vec B_a
+\frac{e}{4\,m^3}\,(\vec p_a\cdot \vec\sigma_a) (\vec B_a \cdot\vec p_a)\bigg]
\nonumber\\&\
+\kappa\sum_{a\neq b}\frac{\alpha}{2\,m^2\, r_{ab}^3}\, \vec\sigma_a\cdot\vec r_{ab}\times\big[e\vec A_a- e\vec A_b \big]\,,\nonumber
\end{align}
where $a$ and $b$ indices refer to the electrons, $\vec A_a = \vec A(\vec r_a)$ and
\begin{align}
e\,\vec A(\vec r) = &\ \frac{e}{4\pi} \vec{\mu}\times\frac{\vec r}{r^3} = -Z\alpha \frac{g}{2M} \vec I\times \frac{\vec r}{r^3}\,.
\end{align}
After the spin averaging $S^i I^j \rightarrow \delta^{ij}\,\vec I\cdot\vec S/3$ with $\vec S = \vec s_1 + \vec s_2$ being the total spin of electrons, it becomes
\begin{align}
H^{(7)}_{ {\rm hfs},A}= &\ \frac{g\,\kappa\,Z\alpha}{4\,m^2\,M} \vec I\cdot\vec S\,\bigg\{
\frac{2\,Z\alpha}{3}\,\frac{1}{r_1^4}
-\frac{4\,\pi}{9\,m}\,p_1^i\,\delta^3(r_1)\,p_1^i \nonumber\\
&\
+\frac{1}{6\,m}\,p_1^i\,\frac{1}{r_1^5}\big(r_1^2\,\delta^{ij}-3\,r_1^i r_1^j\big)\,p_1^j
+\frac{\pi}{3\,m}\,\Delta\,\delta^3(r_1)
\nonumber\\&\
-  \frac43 \alpha\,\frac{\vec r\cdot\vec r_1}{r^3\,r_1^3}\bigg\}
 + (1\leftrightarrow2)\,.
\end{align}
Operator $\Delta\,\delta^3(r_a)$ is transformed into regular $Q_i$ operators from Table \ref{oprsQ1} with the help
of Eq.~(61) of Ref.~\cite{patkos:21:rad}. The second part of $H^{(7)}_{\rm hfs}$ is obtained by expanding
(in $q^2$) the form factors and the vacuum polarization
multiplied by the Fermi contact interaction,
\begin{align}\label{H7b}
 H^{(7)}_{ {\rm hfs},B}
 = &\ \frac{g Z\alpha}{4\,m^3\,M}\, \bigg[F'_1(0)+F'_2(0)-\frac{\alpha}{15\,\pi}\bigg]\,\frac{8\pi}{3}\,\vec I\cdot\vec S
\Delta\,\delta^3(r_1)
 \nonumber\\
&\
+ (1\leftrightarrow2)\,,
\end{align}
where the form-factor slopes are given by
\begin{align}
F'_1(0)+F'_2(0) = \frac{\alpha}{\pi}\bigg[\frac{17}{72} + \frac13\ln\frac{\alpha^{-2}}{2}\bigg]\,.
\end{align}

We now turn to restoring the missing contribution proportional to the electron-nucleus
Dirac $\delta$ function. This is accomplished by evaluating the large-$Z$ limit of
the above formulas. In the $Z$$\to$$\infty$ limit, all effects
of the electron-electron interaction vanish (since they are suppressed by a factor of $1/Z$
as compared to the electron-nucleus interaction) and the result should agree
with the $m\alpha^7$ correction derived for the hydrogen-like ions. This
matching gives us the coefficient at the electron-nucleus
Dirac $\delta$ function. As a result, we obtain an additional first-order
contribution, which reads
\begin{align}\label{eq:eta}
E_{{\rm fo}, B} = &\, \frac{\alpha(\Za)^3g}{4\,\pi\,M}\, \lbr \vec{I}\cdot\vec{S}\rbr\, \pi
 \lbr \big[ \delta^3(r_1)+\delta^3(r_2)\big] \rbr
\nonumber \\ \times  &
\bigg[-\frac{5351}{1350} - \frac{44\pi^2}{27} - \frac{10}{3}\zeta(3) + \frac{896}{27}\ln 2\nonumber\\
&\ + \frac{16}{9}\ln^2 2  - \frac{4882}{135}\ln\alpha - \frac{64}{9}\ln^2\alpha + \frac{256}{9}\ln 2\ln\alpha \bigg]
 \,.
\end{align}
Finally we obtain the total first-order contribution as
\begin{align}
\lbr H^{(7)}_{ {\rm hfs},A}\rbr + \lbr H^{(7)}_{ {\rm hfs},B}\rbr +  E_{{\rm fo}, A} + E_{{\rm fo}, B}
= m\alpha^7\big[ {\cal E}_{\rm fo}(\mathrm{se}) + {\cal E}_{\rm fo}(\mathrm{vp})\big] \,,
\end{align}
where ${\cal E}_{\rm fo}(\mathrm{se})$ and ${\cal E}_{\rm fo}(\mathrm{vp})$ are given by Eqs.~(\ref{fofinQ})
and (\ref{finvp2}), respectively. The details of the derivation will be published elsewhere \cite{pachucki:23:arxiv}.


\end{document}